\documentclass{article}
\usepackage{amsmath,amssymb}
\usepackage{graphicx}

\begin{document}
\title{Magnetism and superconductivity within extended Hubbard model for a dimer.\\Exact results}
\author{B. Grabiec$^1$, S. Krawiec$^2$ and M. Matlak$^2$ {\footnote{Corresponding author. E-mail: matlak@us.edu.pl}}
\\
$^1$Institute of Physics, University of Zielona G\'ora,\\
Prof. Z. Szafrana 4a, 65-516 Zielona G\'ora, Poland\\
$^2$Institute of Physics,University of Silesia,\\ 
Uniwersytecka 4, 40-007 Katowice, Poland}
\maketitle

\begin{abstract}
We consider the extended Hubbard model containing intrasite and intersite
Coloumb interactions. This model we apply to a dimer as an interisting and
nontrivial example of two interacting ions which possesses exact, analytical
solution. We find the eigenvalues $E_{\alpha} $ and eigenvectors $|E_{\alpha}
\rangle $ of the dimer and we represent each part $E_{\alpha} |E_{\alpha}
\rangle \langle E_{\alpha} |$ of the dimer Hamiltonian $(\alpha =1,2,...,16)$
in the second quantizations with the use of the Hubbard and spin operators.
This procedure gives a review of all competitive, intrinsic interactions
together with their exactly calculated coupling constants expressed by the
model parameters. These competitive interactions, deeply hidden in the
orginal form of the extended Hubbard model, make an evidence that the model
is extremely complex because it describes a competition between magnetism
and superconductivity. Among competitive interactions we can find
ferromagnetic and antiferromagnetic interactions (also in the form appearing
in the $t-J$ model), hopping of the Cooper pairs between different dimer
lattice sites (similar as in Kulik-Pedan, Penson-Kolb models), as well as,
intersite Cooper pair interactions. We plot several coupling constants of
these interactions vs Coulomb intrasite interaction $U$ to show that the
competition between them strongly depends on the model parameters. The
thermodynamical activity of each particular interaction, belonging to a
given energy level, depends, however, on the occupation of this level. When
e.g. a given dimer energy level is empty the corresponding intrinsic
interaction belonging to this level is completely passive (irrelevant). From
the presented review of interactions it is evident that the extended Hubbard
model is capable to describe the properties of superconductors with magnetic
ordering (including also high-$T_C$ superconductors) where a strong
competition between magnetism and superconductivity takes place.
\end{abstract}

\section{Introduction}

It is commonly believed that the Hubbard model and its extentions can widely
be used to explain many of physical phenomena observed in quite different
areas of the solid state physics: magnetic and transport properties of
transitions matals, their compounds and alloys, including insulator-metal
transitions (cf e.g. Refs [1]-[13] and papers cited therein),
superconductors (magnetic superconductors (see e.g. Refs [14], [15]), high-$%
T_C$ superconductors (negative $U$ models, cf. e.g. Ref. [16]; $t-J$ based
models (cf Refs [17]-[26])), fluctuating valence systems and heavy fermions
(Anderson-like models, Ref. [27], cf also e.g. Refs [11], [28]), liquid $He^3
$ (cf e.g. Refs [29]-[31]), fullerens (cf e.g. Refs [32]-[34]), etc.. To
demostrate a high degree of complexity of the extended Hubbard model we
consider one-band electronic system described by the Hamiltonian 
\begin{equation}
\begin{array}{ll}
H= & \sum\limits_{i\neq j,\sigma }t_{i,j}c_{i,\sigma }^{+}c_{j,\sigma
}+U\sum\limits_in_{i,\uparrow }n_{i,\downarrow } \\ 
& +\frac 12\sum\limits_{i\neq j,\sigma }J_{ij}^{(1)}n_{i,\sigma }n_{j,\sigma
}+\frac 12\sum\limits_{i\neq j,\sigma }J_{ij}^{(2)}n_{i,\sigma }n_{j,-\sigma
}.
\end{array}
\end{equation}
The indices ($i,j$) enumerate the lattice points $(\mathbf{R_i,R_j)}$, $%
t_{i,j}$ is the hopping integral, $U$ (positive or negative) denotes the
effective intrasite Coloumb interaction, $J^{(1)}$ and $J^{(2)}$ (generally,
not necessary equal) decribe the effective intersite interactions, resulting
from the original Coloumb repulsion modified by the polaronic effects (see
e.g. Ref. [16] for details). The operators $c_{i,\sigma }(c_{i,\sigma }^{+})$
are the electronic annihilation (creation) operators in the lattice site $i$
with spin index $\sigma =\uparrow ,\downarrow $ and $n_{i,\sigma
}=c_{i,\sigma }^{+}c_{i,\sigma }$. The model (1) cannot be solved exactly in
a general case. There is, however, a special but nontrivial case of two
interacting ions (a dimer problem) which possesses exact, analytical
solution. The Hamiltonian of the dimer has the form (see (1))

\begin{equation}
\begin{array}{ll}
H_D= & -t\sum\limits_\sigma (c_{1,\sigma }^{+}c_{2,\sigma }+c_{2,\sigma
}^{+}c_{1,\sigma })+U(n_{1,\uparrow }n_{1,\downarrow }+n_{2,\uparrow
}n_{2,\downarrow }) \\ 
& +J^{(1)}\sum\limits_\sigma n_{1,\sigma }n_{2,\sigma
}+J^{(2)}\sum\limits_\sigma n_{1,\sigma }n_{2,-\sigma }
\end{array}
\end{equation}
where $t=-t_{1,2}=-t_{2,1}$. We can exactly show that $H_D$ incorporates
many competitive interactions, deeply hidden in its orginal form. To bring
them all into light we will follow the method, applied earlier to the
Hubbard dimer alone (see Ref. [35]). The application of this method to the
extended Hubbard model for a dimer seems to be very important because it
gives a review of all intrinsic interactions, possible to appear in this
more general case. According to the Ref. [35] we first find the exact
solution of the dimer problem (see Sect. 2). This solution consists of 16
energy levels $E_{\alpha} $ and corresponding eigenvectors $|E_{\alpha} \rangle $
($\alpha =1,2,...,16$). In the next step we use the equivalent expression
for the dimer Hamiltonian 
\begin{equation}
H_D=\sum\limits_{\alpha =1}^{16}E_{\alpha} P_{\alpha} 
\end{equation}
where $P_{\alpha} =|E_{\alpha} \rangle \langle E_{\alpha} |$. For each $\alpha
=1,2..,16$ the oprator $P_{\alpha} $ can be represented in the second
quantization where we introduce Hubbard and spin operators. When using the
relation (3) we can decompose the dimer Hamiltonian (2) into 10 parts
collecting the terms belonging to the same energy level. Such decomposition
according to the exact dimer energy levels possesses two main adventages.
First, we bring into light all competitive interactions, possible to appear
within the exactly solvable dimer. We find ferromagnetic and
antiferromagnetic interactions, Cooper pair hopping terms (similar as in
Kulik-Pedan, Penson-Kolb models (cf. Refs [36],[37])) with positive and
negative coupling constants, as well as, intersite Cooper pair interactions
(see e.g. Ref [16]). In this way we exactly show that the extended Hubbard
model for a dimer is, in fact, a mixture of different intrinsic interactions
leading to magnetic and superconducting properties of the model which
compete together. Second, the presence of each intrinsic interaction is
stricly ascribed to a given energy level which can be occupied or empty (it
depends on the model parameters and assumed average occupation number of
electrons $n\in [0,4]$). In other words, we can foresee which intrinsic
interaction can be ''thermodynamically'' active or not (it depends on the
position of the chemical potential $\mu $ with respect to a given energy
level). Coupling constants connected with each type of intrinsic interaction
are calculated exactly and several of them we plot as a functions of the
intrasite Coulomb interaction $U$ to visualize the competition between them.
The method applied in this paper is exact in comparison with approximative
approaches (perturbation expansion or canonical transformations) used in the
literature till now  (see Refs [38]-[53] and [11] and [12] for a review).
The knowledge about possible intrinsic interactions within the extended
Hubbard model seems to be very important because this model is widely used
in different areas of the solid state physics. The presence of ferromagnetic
and antiferromagnetic interactions competing with different types of
interactions leading to superconductivity unambigously suggests the use of
the model to magnetic superconductors (cf e.g. Refs [14], [15] and papers
cited therein). There is, however, one important and unsolved problem which
arises from our exact calculations for a dimer. When apply the model to a
real lattice (real materials) we are always forced to make aproximations
which can destroy a delicate harmony of the model leading to the
overestimation or underestimation of some important competitive
interactions. Thus, the problem is how to formulate a resonable
approximation which treats all of them on equal footing. Besides, a high
degree of complexity of the extended Hubbard model, explicitly shown here
for a dimer, can also be a reason why the understanding of pairing in the
high-$T_C$ superconductors is such a difficult problem till now. It seems to
be evident that the $t-J$ model alone (also present in our approach),
introduced in Refs [17]-[26] (see also Ref. [53] for a review), does not
contain enough ingredients to describe high-$T_C$ superconductors because we
additionally have a competition between ferromagnetism and
antiferromagnetism accompanied by a challenge between the hopping of the
Cooper pairs between different lattice sites (similar as in Refs [36], [37])
and intersite Cooper pair interactions.

\section{Intrinsic interactions}

The eigenvalue problem for the dimer Hamiltonian (2) can be solved exactly
when start from the vectors $|n_{1,\uparrow },n_{1,\downarrow
};n_{2,\uparrow },n_{2,\downarrow }\rangle $ $(n_{i,\sigma }=0,1;$ $i=1,2$; $%
\sigma =\uparrow ,\downarrow )$ forming the Fock basis (cf. also Refs [35,
[54], [55]) 
\begin{equation}
\begin{array}{lll}
\begin{array}{l}
\begin{array}{l}
|0\rangle =|0,0;0,0\rangle ,
\end{array}
\\ 
\\ 
\begin{array}{l}
|11\rangle =|1,0;0,0\rangle , \\ 
|12\rangle =|0,1;0,0\rangle , \\ 
|13\rangle =|0,0;1,0\rangle , \\ 
|14\rangle =|0,0;0,1\rangle ,
\end{array}
\end{array}
& 
\begin{array}{l}
|21\rangle =|1,1;0,0\rangle , \\ 
|22\rangle =|1,0;1,0\rangle , \\ 
|23\rangle =|1,0;0,1\rangle , \\ 
|24\rangle =|0,1;1,0\rangle , \\ 
|25\rangle =|0,1;0,1\rangle , \\ 
|26\rangle =|0,0;1,1\rangle ,
\end{array}
& 
\begin{array}{l}
\begin{array}{l}
|31\rangle =|0,1;1,1\rangle , \\ 
|32\rangle =|1,0;1,1\rangle , \\ 
|33\rangle =|1,1;0,1\rangle , \\ 
|34\rangle =|1,1;1,0\rangle ,
\end{array}
\\ 
\\ 
\begin{array}{l}
|4\rangle =|1,1;1,1\rangle .
\end{array}
\end{array}
\end{array}
\end{equation}
The basis vectors (4) have the form $|n\beta \rangle $ where $%
n=\sum_{i,\sigma }n_{i,\sigma }(=0,1,2,3,4)$. The second index $\beta $ (if
necessary) enumerates the vectors belonging to the subspace of a given $n$.
The exact solution of the dimer eigenvalue problem $H_D|E_{\alpha} \rangle
=E_{\alpha} |E_{\alpha} \rangle $ can be obtained with the use of a standard
procedure. We obtain

\[
\begin{array}{ll}
E_1=0; & |E_1\rangle =|0\rangle ,
\end{array}
\]

\[
\begin{array}{ll}
E_{2}=-t; & |E_{2}\rangle =\frac{1}{\sqrt{2}}(|11\rangle +|13\rangle ), \\ 
E_{3}=t; & |E_{3}\rangle =\frac{1}{\sqrt{2}}(|11\rangle -|13\rangle ), \\ 
E_{4}=-t; & |E_{4}\rangle =\frac{1}{\sqrt{2}}(|12\rangle +|14\rangle ), \\ 
E_{5}=t; & |E_{5}\rangle =\frac{1}{\sqrt{2}}(|12\rangle -|14\rangle ),
\end{array}
\]

\begin{equation}
\begin{array}{ll}
E_{6}=J^{(2)}; & |E_{6}\rangle =\frac 1{\sqrt{2}}(|23\rangle +|24\rangle ),
\\ 
E_{7}=U; & |E_{7}\rangle =\frac 1{\sqrt{2}}(|21\rangle -|26\rangle ), \\ 
E_{8}=C+\frac{U+J^{(2)}}{2}; & |E_{8}\rangle =a_1(|21\rangle +|26\rangle
)-a_2(|23\rangle -|24\rangle ), \\ 
E_{9}=-C+\frac{U+J^{(2)}}{2}; & |E_{9}\rangle =a_2(|21\rangle +|26\rangle
)+a_1(|23\rangle -|24\rangle ), \\ 
E_{10}=J^{(1)}; & |E_{10}\rangle =|22\rangle , \\ 
E_{11}=J^{(1)}; & |E_{11}\rangle =|25\rangle ,
\end{array}
\end{equation}

\[
\begin{array}{ll}
E_{12}=t+U+J^{(1)}+J^{(2)}; & |E_{12}\rangle =\frac 1{\sqrt{2}}(|31\rangle
+|33\rangle ), \\ 
E_{13}=-t+U+J^{(1)}+J^{(2)}; & |E_{13}\rangle =\frac 1{\sqrt{2}}(|31\rangle
-|33\rangle ), \\ 
E_{14}=t+U+J^{(1)}+J^{(2)}; & |E_{14}\rangle =\frac 1{\sqrt{2}}(|32\rangle
+|34\rangle ), \\ 
E_{15}=-t+U+J^{(1)}+J^{(2)}; & |E_{15}\rangle =\frac 1{\sqrt{2}}(|32\rangle
-|34\rangle ),
\end{array}
\]

\[
\begin{array}{ll}
E_{16}=2(U+J^{(1)}+J^{(2)}); & |E_{16}\rangle =|4\rangle
\end{array}
\]
where

\begin{equation}
C=\sqrt{{\left( \frac{U-J^{(2)}}2\right) }^2+4t^2},
\end{equation}

\begin{equation}
a_1=\frac 12\sqrt{1+\frac{(U-J^{(2)})}{2C}},
\end{equation}

\begin{equation}
a_2=\frac 12\sqrt{1-\frac{(U-J^{(2)})}{2C}}.
\end{equation}

Lets us introduce the Hubbard operators

\begin{equation}
a_{i,\sigma }=c_{i,\sigma }(1-n_{i,-\sigma }),
\end{equation}
\begin{equation}
b_{i,\sigma }=c_{i,\sigma }n_{i,-\sigma }
\end{equation}
\ \newline
and spin operators 
\begin{equation}
S_i^z=\frac 12(n_{i,\uparrow }-n_{i,\downarrow })=\frac 12(n_{i,\uparrow
}^a-n_{i,\downarrow }^a),
\end{equation}
\begin{equation}
S_i^{+}=c_{i,\uparrow }^{+}c_{i,\downarrow }=a_{i,\uparrow
}^{+}a_{i,\downarrow },
\end{equation}
\begin{equation}
S_i^{-}=c_{i,\downarrow }^{+}c_{i,\uparrow }=a_{i,\downarrow
}^{+}a_{i,\uparrow }
\end{equation}
where $n_{i,\sigma }^a=a_{i,\sigma }^{+}a_{i,\sigma }$ $(i=1,2;$ $\sigma
=\uparrow ,\downarrow )$.

To obtain all intrinsic interactions, deeply hidden in the original form of
the extended Hubbard model for a dimer (2) we apply the equivalent form the
dimer Hamiltonian (3). Each product $E_{\alpha} P_{\alpha} $ in the formula (3)
where we insert $E_{\alpha} $ and $|E_{\alpha} \rangle $ from the formulae (5)
can be represented as a linear combination of the basis vectors (4) which,
in turn, with the use of the Hubbard and spin operators (9)-(13) can be
rewritten in the second quantization (see Refs. [35], [54], [55]). We
collect all the terms together which correspond to the same energy level (we
take into account the degeneration of the levels as e.g. $E_4=E_2$, $E_5=E_3$%
, $E_{11}=E_{10}$, $E_{14}=E_{12}$, $E_{15}=E_{13}$). In this way we can
split the Hamiltonian (2) written in the form of the formula (3) into 10
terms, corresponding to 10 different dimer energy levels (see (5)). We obtain

\begin{equation}
H_D=\sum_{i=1}^{10} H_d^{(i)}
\end{equation}

where

\begin{eqnarray}
H_D^{(1)}&=&E_{2}P_{2}+E_{4}P_{4}=-\frac t2[n_1^a(1-n_2^a-\frac{n_2^b}%
2)+n_2^a(1-n_1^a-\frac{n_1^b}2)]  \nonumber \\
&&-\frac t2\sum_\sigma [a_{1,\sigma }^{+}a_{2,\sigma }+a_{2,\sigma
}^{+}a_{1,\sigma }],
\end{eqnarray}

\begin{eqnarray}
H_D^{(2)}&=&E_{3}P_{3}+E_{5}P_{5}=\frac t2[n_1^a(1-n_2^a-\frac{n_2^b}%
2)+n_2^a(1-n_1^a-\frac{n_1^b}2)]  \nonumber \\
&&-\frac t2\sum_\sigma [a_{1,\sigma }^{+}a_{2,\sigma }+a_{2,\sigma
}^{+}a_{1,\sigma }],
\end{eqnarray}

\begin{equation}
H_D^{(3)}=E_6P_6=-J^{(2)}[{S_1}^z\cdot {S_2}^z-\frac{n_1^an_2^a}4]+\frac{%
J^{(2)}}2\left( {S_1}^{+}\cdot {S_2}^{-}+{S_1}^{-}\cdot {S_2}^{+}\right)
\end{equation}

\begin{eqnarray}
H_D^{(4)}&=&E_7P_7=\frac U4[n_1^b(1-n_2^a-\frac{n_2^b}2)+n_2^b(1-n_1^a-\frac{%
n_1^b}2)] \nonumber \\
&&-\frac U2[b_{1,\uparrow }^{+}a_{1,\downarrow }^{+}a_{2,\downarrow
}b_{2,\uparrow }+b_{2,\uparrow }^{+}a_{2,\downarrow }^{+}a_{1,\downarrow
}b_{1,\uparrow }],
\end{eqnarray}

\begin{eqnarray}
H_D^{(5)} &=&E_8P_8=\left\{ -\frac{J^{(2)}}2+[\frac{J^{(2)}(U-J^{(2)})}{4C}-%
\frac{2t^2}C]\right\} [{\mathbf{\overrightarrow{S_1}\cdot \overrightarrow{S_2%
}}}-\frac{n_1^an_2^a}4]  \nonumber \\
&&+\frac 14[U+C+\frac{(U^2-(J^{(2)})^2)}{4C}][b_{1,\uparrow
}^{+}a_{1,\downarrow }^{+}a_{2,\downarrow }b_{2,\uparrow }+b_{2,\uparrow
}^{+}a_{2,\downarrow }^{+}a_{1,\downarrow }b_{1,\uparrow }]  \nonumber \\
&&+\left\{ \frac U8+\frac 18[C+\frac{(U^2-(J^{(2)})^2)}{4C}]\right\}
[n_1^b(1-n_2^a-\frac{n_2^b}2)+n_2^b(1-n_1^a-\frac{n_1^b}2)]  \nonumber \\
&&+\left\{ -\frac t2-\frac{t(U+J^{(2)})}{4C}\right\} \sum_\sigma
\sum_{i=1}^2[a_{i,\sigma }^{+}b_{\overline{i},\sigma }+b_{i,\sigma }^{+}a_{%
\overline{i},\sigma }],
\end{eqnarray}

\begin{eqnarray}
H_D^{(6)} &=&E_9P_9=\left\{ -\frac{J^{(2)}}2-[\frac{J^{(2)}(U-J^{(2)})}{4C}-%
\frac{2t^2}C]\right\} [{\mathbf{\overrightarrow{S_1}\cdot \overrightarrow{S_2%
}}}-\frac{n_1^an_2^a}4]  \nonumber \\
&&+\frac 14[U-C-\frac{(U^2-(J^{(2)})^2)}{4C}][b_{1,\uparrow
}^{+}a_{1,\downarrow }^{+}a_{2,\downarrow }b_{2,\uparrow }+b_{2,\uparrow
}^{+}a_{2,\downarrow }^{+}a_{1,\downarrow }b_{1,\uparrow }]  \nonumber \\
&&+\left\{ \frac U8-\frac 18[C+\frac{(U^2-(J^{(2)})^2)}{4C}]\right\}
[n_1^b(1-n_2^a-\frac{n_2^b}2)+n_2^b(1-n_1^a-\frac{n_1^b}2)]  \nonumber \\
&&+\left\{ -\frac t2+\frac{t(U+J^{(2)})}{4C}\right\} \sum_\sigma
\sum_{i=1}^2[a_{i,\sigma }^{+}b_{\overline{i},\sigma }+b_{i,\sigma }^{+}a_{%
\overline{i},\sigma }],
\end{eqnarray}

\begin{equation}
H_D^{(7)}=E_{10}P_{10}+E_{11}P_{11}=2J^{(1)}[{S_1}^z\cdot{S_2}^z+\frac{%
n_1^an_2^a}4]
\end{equation}

\begin{eqnarray}
H_D^{(8)} &=&E_{12}P_{12}+E_{14}P_{14}=\frac{(t+U+J^{(1)}+J^{(2)})}%
4[n_1^an_2^b+n_2^an_1^b]  \nonumber \\
&&-\frac{(t+U+J^{(1)}+J^{(2)})}2\sum_\sigma [b_{1,\sigma }^{+}b_{2,\sigma
}+b_{2,\sigma }^{+}b_{1,\sigma }],
\end{eqnarray}

\begin{eqnarray}
H_D^{(9)} &=&E_{13}P_{13}+E_{15}P_{15}=\frac{(-t+U+J^{(1)}+J^{(2)})}%
4[n_1^an_2^b+n_2^an_1^b]  \nonumber \\
&&+\frac{(-t+U+J^{(1)}+J^{(2)})}2\sum_\sigma [b_{1,\sigma }^{+}b_{2,\sigma
}+b_{2,\sigma }^{+}b_{1,\sigma }],
\end{eqnarray}

\begin{equation}
H_D^{(10)}=E_{16}P_{16}=\frac{(U+J^{(1)}+J^{(2)})}2n_1^bn_2^b,
\end{equation}

and $n_i^{a,b}=\sum_\sigma n_{i,\sigma }^{a,b}$, $n_{i,\sigma
}^b=b_{i,\sigma }^{+}b_{i,\sigma }=n_{i,\sigma }n_{i,-\sigma }$ $(i=1,2)$, $%
\overline{i}=1$ when $i=2$ and $\overline{i}=2$ when $i=1$. As a test for
the correctness of the decomposition (14) we can sum up all the terms
(15)-(24) and we obtain the dimer Hamiltonian in its original form (2). The
decomposition of the dimer Hamiltonian (14) according to the dimer energy
levels where each part is given by the expressions (15)-(24) has, however,
very important advantages. First of all it vizualizes in an explicit and
exact way all possible intrinsic interactions with corresponding coupling
constants. Second, due to the decomposition according to the dimer energy
levels we can easily see which interaction can be active or passive. It is
simply connected with the occupation of a given level. If a dimer level is
occupied by the electrons the intrinsic interaction ascribed to this level
will be active, otherwise passive (it can be neglected). The occupation, in
turn, depends on the model parameters, temperature and the assumed average
number of electrons $n=\sum_{i,\sigma }\langle n_{i,\sigma }\rangle $ which
determines the position of the chemical potential and decides which energy
levels will be occupied or empty.

\begin{figure}[h]
\centering
 \includegraphics[angle=-90,scale=0.35]{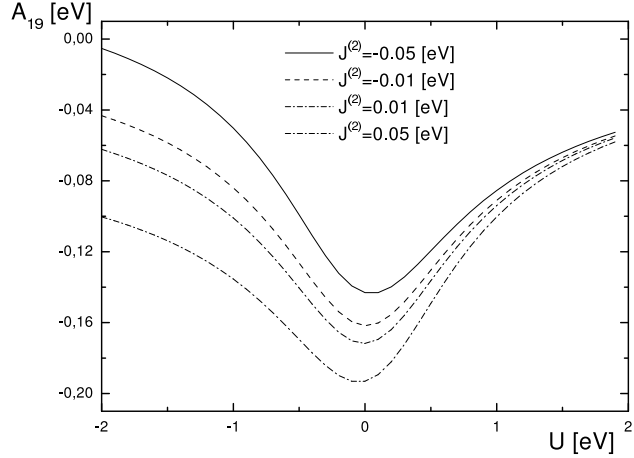}
 \caption{Plot of the coupling constant $A_{19}$ (the coefficient at the first
term in the formula (19)) vs Coulomb interaction $U$ for several values of $%
J^{(2)}$. The parameter $t=1/6$ $eV$ for all Figs.}
\end{figure}

\begin{figure}[h]
\centering
 \includegraphics[angle=-90,scale=0.35]{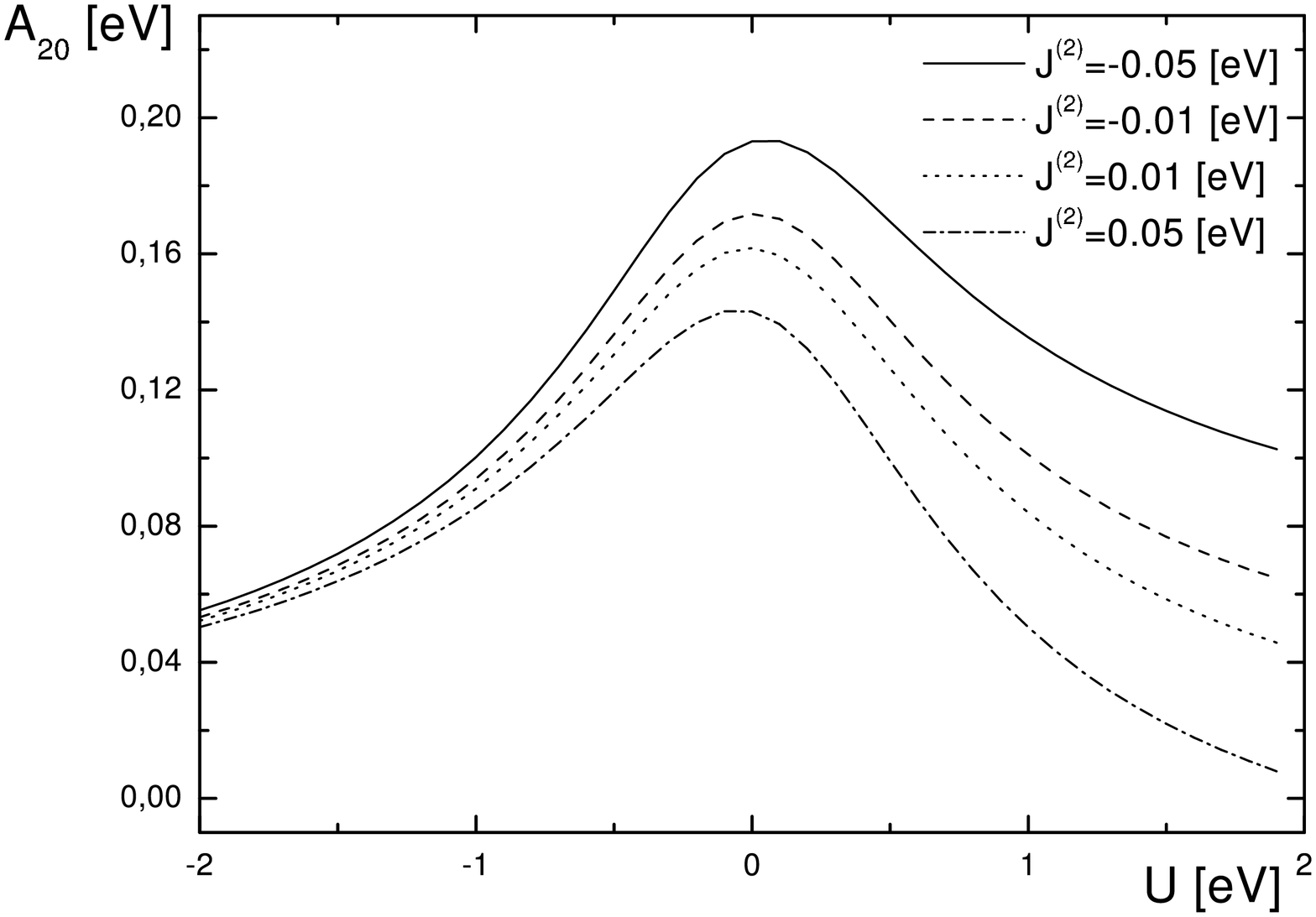}
 \caption{Plot of the coupling constant $A_{20}$ (the coefficient at the first
term in the formula (20)) vs Coulomb interaction $U$ for several values of $%
J^{(2)}$.}
\end{figure}

The most important intrinsic interactions (leading to magnetism or
superconductivity), presented in the expressions (15)-(24) can be devided
into two classes. First of them belongs to magnetic interactions
(ferromagnetic, antiferromagnetic - it depends on the sign of the parametrs $%
J^{(1)}$, $J^{(2)}$). Such interactions are present in the formulae (17) and
(21) and describe Ising type interactions with coupling constants generated
by $J^{(1)}$ and $J^{(2)}$. The formuale (17) contains also the transverse
interaction between spins (see also later). The Heisenberg type magnetic
interactions can be seen in the first terms of the formulae (19) and (20),
generated by a more complex coupling constants, expressed by the model
parameters $J^{(2)}$, $U$ and $t$. It is interesting to note that when $%
J^{(2)}=0$ the first term in the formulae (19) and (20) describes
ferromagnetic or antiferromagnetic interactions (it depends on the sign of $%
U $), similar to well-known $t-J$ model because the coefficient $\frac{2t^2}%
C\approx \frac{4t^2}{\mid U\mid }$ for large $\mid U\mid $ (see the formula
(6) when $J^{(2)}=0$). The first terms in the expressions (19) and (20) can
also be considered as generalized $t-J$ interactions, valid in a more
general case of the extended Hubbard model. The coefficients $A_{19}=-\frac{%
J^{(2)}}2+[\frac{J^{(2)}(U-J^{(2)})}{4C}-\frac{2t^2}C]$ (the coupling
constant at the first term in (19)) and $A_{20}=-\frac{J^{(2)}}2-[\frac{%
J^{(2)}(U-J^{(2)})}{4C}-\frac{2t^2}C]$ (the coupling constant at the first
term in (20)) are depicted in Fig.1 and Fig.2, respectively, as functions of
the intrasite Coulomb interaction $U$ for serveral values of the parameter $%
J^{(2)}$. We see that $A_{19}$ is negative and $A_{20}$ is positive. Thus,
the first terms in (19) and (20) describe competitive ferromagnetic and
antiferromagnetic Heisenberg interactions.

\begin{figure}[h]
\centering
 \includegraphics[angle=-90, scale=0.35]{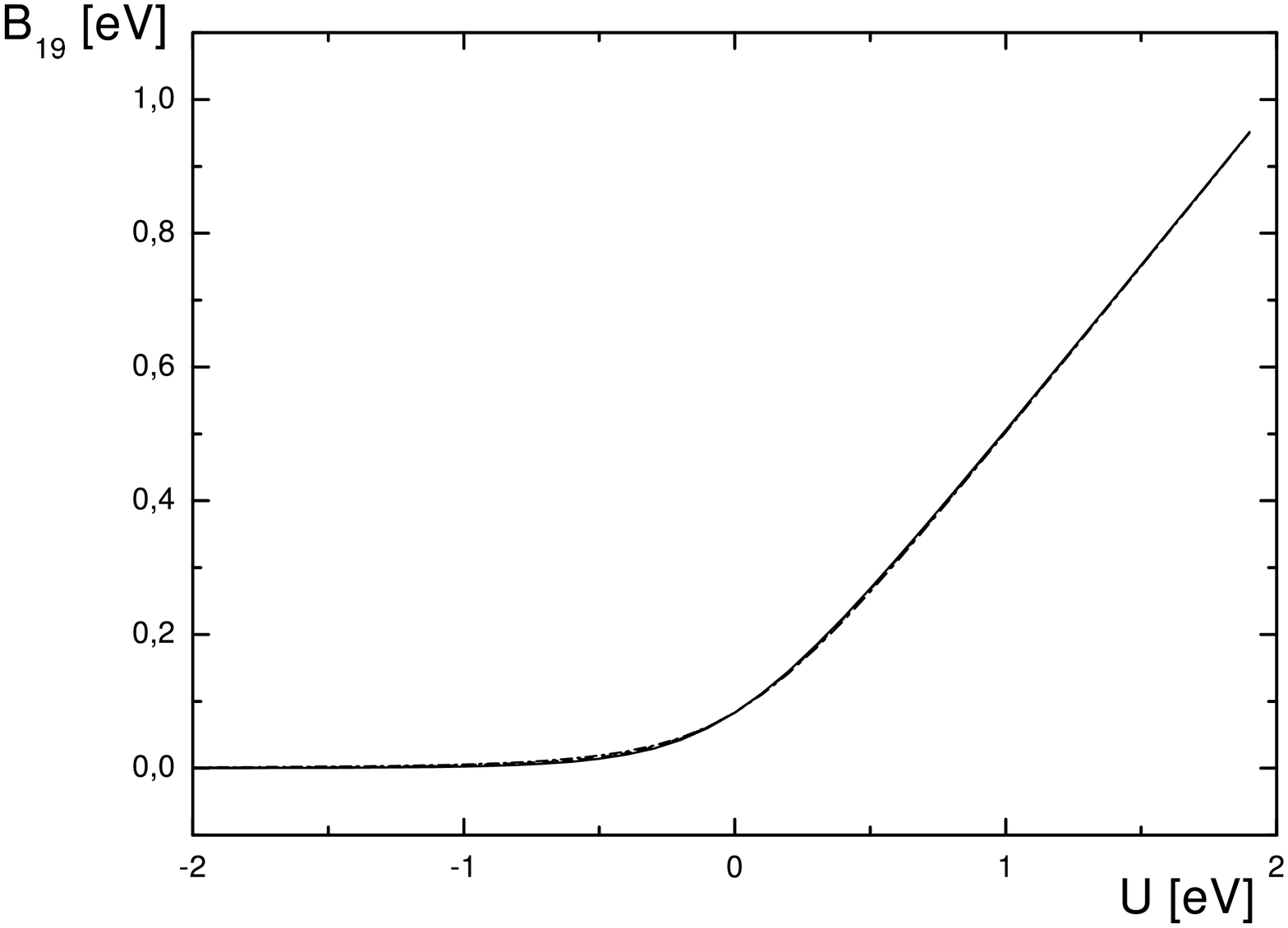}
 \caption{Plot of the coupling constant $B_{19}$ (the coefficient at the second
term in the formula (19)) vs Coulomb interaction $U$. The values of $J^{(2)}$%
(as in Figs 1 and 2) have no significant influence on the plot in this scale.}
\end{figure}

\begin{figure}[h]
\centering
 \includegraphics[angle=-90, scale=0.35]{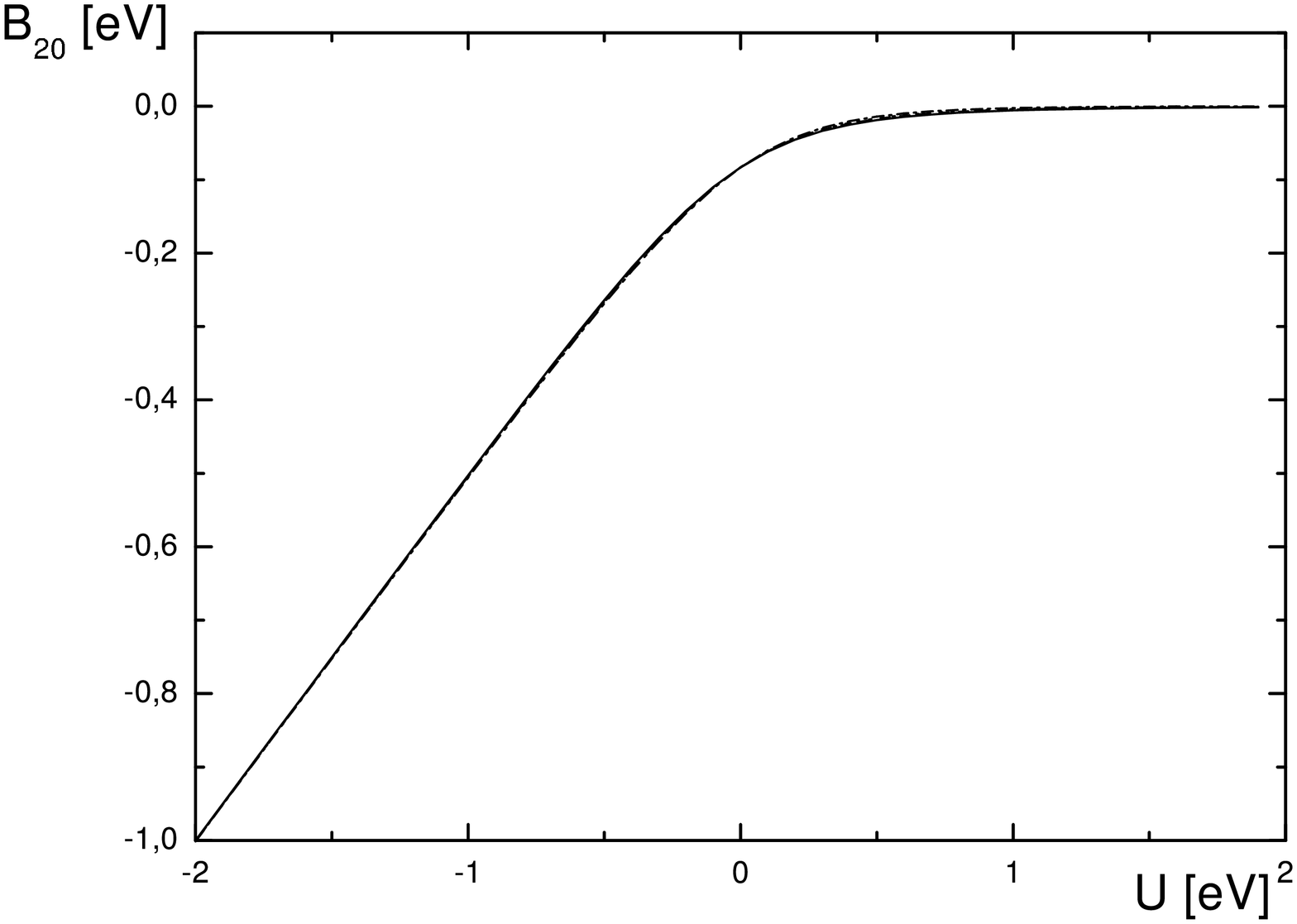}
 \caption{Fig.4. Plot of the coupling constant $B_{20}$ (the coefficient at the second
term in the formula (20)) vs Coulomb interaction $U$. The values of $J^{(2)}$%
(as in Figs 1 and 2) have no significant influence on the plot in this scale.}
\end{figure}

Let us note that the terms like $b_{1,\uparrow }^{+}a_{1,\downarrow
}^{+}a_{2,\downarrow }b_{2,\uparrow }=c_{1,\uparrow }^{+}c_{1,\downarrow
}^{+}c_{2,\downarrow }c_{2,\uparrow }$ ($b_{2,\uparrow }^{+}a_{2,\downarrow
}^{+}a_{1,\downarrow }b_{1,\uparrow }=c_{2,\uparrow }^{+}c_{2,\downarrow
}^{+}c_{1,\downarrow }c_{1,\uparrow }$) describe the hopping of the Cooper
pairs (see the formulae (9) and (10)). Such a competitive hopping is present
in the second terms in (18), (19) and (20). The plots of the corresponding
coupling constants $B_{19}=\frac 14[U+C+\frac{(U^2-(J^{(2)})^2)}{4C}]$ (the
second term in the formulae (19)) and $B_{20}=\frac 14[U-C-\frac{%
(U^2-(J^{(2)})^2)}{4C}]$ (the second term in the formulae (20)) are depicted
in Fig.3 and Fig.4, respectively. The coupling constant $B_{19}$ is positive
and $B_{20}$ negative. Thus, the second terms in the formulae (19) and (20)
describe the hopping of the Cooper pairs with positive and negative coupling
constants. Such terms are present in the Kulik-Pedan, Penson-Kolb
superconductivity models (cf. Refs [36], [37]).

The transversal products ${S_1}^{+}{S_2}^{-}$ (${S_1}^{-}{S_2}^{+}$) are
present in the second term in the formulae (17), as well as, in the scalar
products of spin operators ${{\vec{S}}_1\cdot }{{\vec{S}}_2}={S_1}^z{S_2}%
^z+\frac 12({S_1^{+}S}_2^{-}+{S_1^{-}S_2^{+}})$ in the first terms of the
formulae (19) and (20). When use the second quantization representation we
obtain ${S_1^{+}S_2^{-}}=-c_{1,\uparrow }^{+}c_{2,\downarrow
}^{+}c_{1,\downarrow }c_{2,\uparrow }$ (${S_1^{-}S_2}^{+}=-c_{2,\uparrow
}^{+}c_{1,\downarrow }^{+}c_{2,\downarrow }c_{1,\uparrow }$) and we can see
that such products of spin operators describe the interaction of the
intersite Cooper pairs because we can write ${S_1^{+}S_2^{-}=-}%
d_{2,1}^{+}d_{1,2}$ (${S_1^{-}S_2}^{+}=-d_{1,2}^{+}d_{2,1}$) where $%
d_{1,2}=c_{1,\downarrow }c_{2,\uparrow }$ ($d_{2,1}=c_{2,\downarrow
}c_{1,\uparrow }$). Thus, the second term in the formulae (17) and the
transversal parts in the first terms of the formulae (19) and (20) describe,
in fact, the competitive intersite Cooper pair interactions leading to
superconductivity (cf. e.g. Ref [16] and papers cited therein). From the
other side the application of the resonating valence bond approach (cf. Refs
[17]-[26]) allows to rewrite the terms like $({\mathbf{\overrightarrow{S_1}%
\cdot \overrightarrow{S_2}}}-\frac{n_1^an_2^a}4$), present in the first
terms of the expressions (19) and (20), when introducing the following
operators $f_{2,1}=$ $\frac 1{\sqrt{2}}(c_{2,\downarrow }c_{1,\uparrow
}-c_{2,\uparrow }c_{1,\downarrow })$ and $f_{2,1}^{+}=$ $\frac 1{\sqrt{2}%
}(c_{1,\uparrow }^{+}c_{2,\downarrow }^{+}-c_{1,\downarrow
}^{+}c_{2,\uparrow }^{+})$. For example, in the case of half-filling (cf.
Ref. [21]) we can write $({\mathbf{\overrightarrow{S_1}\cdot \overrightarrow{%
S_2}}}-\frac 14)=-$ $f_{2,1}^{+}f_{2,1}$.

\section{Conclusions}

Using the exact solution of the extended Hubbard model for a dimer we have
reviewed the main, important interactions leading to magnetic and
superconducting properties of the model. In this way we have demonstrated
that the relative simplicity of the extended Hubbard model is only an
illusion. We have shown in an exact way that the magnetic properties
(ferromagnetism, antiferromagnetism) compete together. There is also
additionally a challenge between two types of Cooper pair interactions
(Cooper pair hopping and intersite Cooper pair interactions). Thus, at last,
magnetic and superconducting tendencies of the model compete together. The
final, thermodynamical properties of the extended Hubbard model are then a
result of this competition. Which interaction wins in such a rally it
strongly depends on temperature (there is possible a change of the leader
with varying temperature), model parameters and on the value of the assumed
average number of electrons, determining the chemical potential of the
system. Such indications can certainly be helpful when use the extended
Hubbard model for real lattices. We have also indicated that the extended
single-band Hubbard model is, in fact, capable to describe properties of
such superconductors where the competition between magnetism and
superconductivity can explain many astonishing properties of these materials
(cf e.g. Refs [14]-[25]), including also high-$T_C$ materials and recently
synthetized $CoO_2$ layer based systems (cf Ref. [26] and papers cited
therein).

The nonperturbative formalism presented in this paper is, in principle,
applicable to more complicated clusters consisting e.g. of one central ion
and $z$ its nearest neighbours. We, however, know (see e.g. Refs [56]-[59]
and papers cited therein) that the mathematical problems in this case
exponentially increase with the size of the considered cluster. Another
examples of a cluster approach to Hubbard-like models are given in Refs
[60]-[65].

\end{document}